\documentstyle[12pt]{article}
\begin{document}
 
\begin{titlepage}
\pagestyle{empty}
\baselineskip=21pt
\begin{center}
{\large{\bf In Defence of the `Tunneling' Wave Function of the Universe}}
\end{center}
\vskip .1in
\begin{center}
{Jaume Garriga}

{\it Center For Theoretical Physics, Laboratory for Nuclear Science}
{\it and Department of Physics, Massachusetts Institute of Technology}
{\it Cambridge, MA 02139}

{\it and}

{\it IFAE, Departament de F\'\i sica, Universitat Aut\`onoma de Barcelona,}

{\it 08193 Bellaterra (Barcelona), Spain.}

 {and} 

{Alexander Vilenkin}

{\it Institute of Cosmology, Department of Physics and Astronomy}

{\it Tufts University, Medford, MA 02155, USA}

\vskip .1in

\end{center}
\vskip .5in
\centerline{ {\bf Abstract} }
\baselineskip=18pt    

The tunneling approach to the wave function of the universe has been
recently criticized by Bousso and Hawking who claim that it predicts a
catastrophic instability of de Sitter space with respect to pair
production of black holes.  We show that this claim
is unfounded.  First, we argue that
different horizon size regions in de Sitter space
cannot be treated as independently 
created, as they contend. And second, the WKB tunneling wave function is
not simply the `inverse' of the Hartle-Hawking one, except in very
special cases. Applied to the related problem of pair production of
massive particles, we argue that the tunneling wave function leads to a
small constant production rate, and not to a catastrophy as Bousso and
Hawking's argument would suggest.

\noindent

\end{titlepage}
\newpage
\baselineskip=18pt

\noindent{\bf 1.\quad  Introduction}

\smallskip

The tunneling proposal for the wave function of the universe has been
recently criticized by Bousso and Hawking \cite{1} who argue that it
predicts a catastrophic instability of de Sitter space with respect to
production of black hole pairs.  Here it will be shown that their
argument is incorrect.

Let us first review the Bousso-Hawking argument.  First, they
estimate the pair production rate for black holes in an
inflating universe as 
\begin{equation}
\Gamma \propto \exp [-S_{BH}+S_0],
\label{1}
\end{equation}
where $S_0$ and $S_{BH}$ are Euclidean actions for de Sitter and
(degenerate) Sczwarzschild-de Sitter instantons, respectively.  The
Schwarzschild-de Sitter instanton, which has the topology of $S_2
\times S_2$, is interpreted \cite{GP} as describing
nucleation of two black holes whose horizon radii are equal to 
the radius of the de Sitter horizon, $H^{-1}$.  A simple calculation
gives (in Planck units)  
\begin{equation}
\Gamma \propto \exp (-\pi/3H^2).
\label{2}
\end{equation}
For $H\ll 1$, black hole nucleation is exponentially suppressed.
Evaluation of the tunneling rates using instanton techniques is a
standard practice in the literature, and results similar to (\ref{1}),
(\ref{2}) for black hole nucleation have been previously derived by
Mellor and Moss \cite{2} and others \cite{3}.  We have no objections
against these results \cite{4}.

Now, Bousso and Hawking suggest that instead of using the standard
instanton approach, the nucleation rate can be calculated directly
from the wave function of the universe.  They argue that each horizon
volume in an inflating universe can be regarded as having been
nucleated independently of other horizon regions.  The rate of black
hole production can then be found as
\begin{equation}
\Gamma\propto P_{BH}/P_0,
\label{3}
\end{equation}
where $P_{BH}$ is the probability of quantum nucleation for a
horizon-size universe containing a pair of black holes, and $P_0$ is
the same without black holes.

The nucleation probability in quantum cosmology is sensitive to the
choice of boundary conditions for the wave function of the universe.
With Hartle-Hawking boundary conditions, the wave function is given by
the integral \cite{5}
\begin{equation}
\psi_{HH}=\int e^{-S},
\label{4}
\end{equation}
where $S$ is the Euclidean action and the integration is taken over
compact Euclidean geometries and matter fields with a specified field
configuration at the boundary.  The boundary configuration is chosen
to be the 3-geometry at the moment of nucleation.  For a de Sitter
universe this is the maximal $S_3$, while for black holes in de Sitter
it is $S_1\times S_2$.  The probability is found from $P\propto
|\psi|^2$.  Assuming that the severe divergence problems of the
integral (\ref{4}) are somehow resolved, one can expect that the
dominant contribution to (\ref{4}) is given by the instantons, 
\begin{equation}
\psi_{HH}\propto e^{-S},
\label{5}
\end{equation}
so that $P\propto e^{-2S}$, where $2S$ is the corresponding instanton
action.  Then Eq. (\ref{3}) reduces to (\ref{1}) and (\ref{2}).  This
completes the quantum-cosmological derivation of the black hole pair
production rate.

Turning to the tunneling wave function, Bousso and Hawking argue that,
since it is suppressed rather than amplified under the barrier, this
wave function should be given by 
\begin{equation}
\hbox{``}\psi_T\hbox{''}\propto e^{+S}.
\label{6}
\end{equation}
Then the sign of the exponent in (\ref{1}) is changed to the opposite,
and Eq. (\ref{2}) is replaced by $\Gamma\propto\exp (+\pi/3H^2)$.
This shows no suppression of black hole production, indicating an
instability of de Sitter space.

We have two objections to this argument.  First, we disagree that
horizon volumes in an inflating universe can be treated as nucleating
independently.  Second, the tunneling wave function is not given by
Eq. (\ref{6}), except in some very special cases.

In the next Section we explain why we believe that different horizon
volumes cannot be regarded as nucleating independently.  In Section 3
we discuss the tunneling boundary condition for $\psi$ and explain why
it is very difficult to implement in the problem of black hole
nucleation.  Fortunately, the tunneling wave function can be found in
a very similar problem of massive particle production during
inflation.  The Bousso-Hawking argument can be applied to this problem
without change.  If the argument were true, the tunneling wave
function would predict a catastrophic instability of de Sitter space
with respect to particle production.  However, we show in Section 4
that it actually predicts a small, constant pair production rate, in
agreement with instanton calculations.

\bigskip

\noindent{\bf 2.\quad  Black hole nucleation {\it vs.} nucleation of
universes with black holes}

\smallskip 

To see why different horizon volumes cannot be regarded as nucleated
independently, consider a generic case of inflation driven by a scalar
field $\phi$ with a slowly-varying potential $V(\phi)$, disregarding
for the time being the process of black hole production.  The field
$\phi$ is approximately constant on the horizon scale, and nucleation
of horizon-size universes with different values of $\phi$ can be
approximately described by de Sitter instantons.  The corresponding
probability is
\begin{equation}
P(\phi)\propto\exp [\pm \pi/H^2(\phi)],
\label{7}
\end{equation}
where $H^2(\phi)=8\pi V(\phi)/3$ and 
the upper and lower signs are for the Hartle-Hawking and
tunneling wave functions, respectively \cite{5,6,7}.  (This is one
example when the two wave functions do differ mainly by the sign in
the exponent).  Suppose an observer occupies a horizon-size region
with $\phi=\phi_0$ and wants to know the probability distribution for
$\phi$ in the adjacent regions.  If different horizon volumes were
nucleated independently, then this distribution would be given by
(\ref{7}) and would be independent of $\phi_0$.  This would imply
rapid variation of the field $\phi$ from one horizon region to
another.  Of course, the observer will not be able to see this
variation as long as inflation continues.  But after inflation, all
these horizon volumes will eventually come within the observer's
horizon, and rapid variation of $\phi$ will manifest itself in large
density fluctuations on all scales.  This conclusion is in a sharp
disagreement with the standard analysis of density fluctuations in the
inflationary universe \cite{8}.

The evolution of the field $\phi$ during inflation is determined by
the quantum state of the field.  It is usually assumed that this state
is close to the de Sitter-invariant (Bunch-Davies) vacuum.  Then, it
has been shown \cite{9} that the evolution of $\phi$ can be pictured
as a diffusion process, resulting in strong correlations between the
values of $\phi$ in adjacent horizon regions.  As a consequence,
density fluctuations are small on sufficiently small scales.  It is
possible that some highly excited state of the field would result in
evolution suggested by the random nucleation picture.  Hence, the
quantum-cosmological problem of finding the probability for the
universe to nucleate with a given value of $\phi$ is not the same as
the problem of finding the probability distribution for $\phi$ within
the universe.  Quantum cosmology can be said to determine the latter
distribution only in the sense that it determines the initial quantum
state of $\phi$ at the nucleation of the universe.

Quite similarly, the probability of black hole pair production in an
inflating universe is determined by the quantum state of the
gravitational field.  It is not related in any simple way to the
probability for the universe to nucleate with a pair of black holes.
Eq. (\ref{3}) with the Hartle-Hawking wave function $\psi_{HH}$
correctly gives the leading exponent of the pair production rate
$\Gamma$ only because $\psi_{HH}$ is defined in terms of the same
Euclidean path integrals used in the instanton evaluation of $\Gamma$.

\bigskip

\noindent{\bf 3.\quad  The tunneling wave function}

\smallskip 

The wave function of the universe $\psi$ is defined on superspace,
which is the space of all 3-geometries (and matter field
configurations).  It satisfies the Wheeler-DeWitt equation
\begin{equation}
{\cal H}\psi=0,
\label{8}
\end{equation}
where ${\cal H}$ is a differential operator on superspace.  The
tunneling boundary condition, as defined in Refs. \cite{7,7'}, requires
that $\psi$ includes only outgoing waves at singular boundaries of
superspace.  The physical meaning of this condition is that the
universe originates in a non-singular way (at the regular boundary),
but may end at a singularity.  The regular boundary of superspace
includes all singular 3-geometries which can be obtained as
non-degenerate slices of smooth Euclidean 4-geometries.

The word ``non-degenerate'' means that the presence and the nature of
the singularity are stable with respect to small perturbations of the
slicing.  To give an example of a degenerate slicing, consider a torus
lying on a horizontal surface and imagine slicing it with horizontal
planes.  The slice at the bottom is a circle.  But if the torus is
slightly tilted, the circular slice disappears, and the bottom slice
is an isolated point.  The circular slice is degenerate in the sense
that it is present only for a very special slicing.  A rigorous
definition of non-degenerate slices is given in Morse theory
\cite{10,11}.  It can be shown that singularities on non-degenerate
slices always occur at isolated points.

In addition to the tunneling boundary condition, one has to impose
some kind of regularity condition, e.g., $|\psi|<\infty$.

We do not know how to solve the Wheeler-DeWitt equation (\ref{8}),
except in simple models of high symmetry, when the infinite number of
degrees of freedom of the universe can be reduced to one or two, with
the remaining degrees of freedom being treated as small perturbations.
For example, the nucleation of an inflating universe (without black
holes) can be described by treating the radius of the universe, $a$,
as a non-perturbative variable and deviations from spherical symmetry
as perturbations.  The regular boundary of superspace is then at
$a=0$, with finite amplitudes of the perturbations.  

To find the tunneling wave function for a universe with a pair of
black holes, one needs to construct an appropriate minisuperspace
model.  The $S_2\times S_2$ instanton and the symmetry of the problem
suggest a minisuperspace consisting of 3-geometries having topology
$S_1\times S_2$ with the radius of the sphere $S_2$ fixed at $H^{-1}$.
This model would have the radius $r$ of the circle $S_1$ as its only
variable and would certainly be simple enough to solve.  However, the
minisuperspace boundary at $r=0$ cannot be regarded as regular
boundary.  This is clear from the fact that this singular geometry has
singularities on a 2-dimensional surface ($S_2$), rather than at an
isolated point \cite{12}.  A more complicated model, where the radii
of $S_1$ and $S_2$ are both allowed to vary, has a similar problem.
The tunneling boundary condition cannot be implemented in such models.
Moreover, it appears that such models are not suitable as
minisuperspaces, since they include configurations which are unstable
with respect to small perturbations.  Extensions of $S_1\times S_2$
models that would not have this problem are bound to be less symmetric
and far more difficult to solve.  We note that even if we managed to
construct a suitable minisuperspace model and solve for the tunneling
wave function $\psi_T$, this result would be relevant only for the
problem of the nucleation of a universe with a pair of black holes,
and not for the rate of black hole production during inflation. 

\bigskip

\noindent{\bf 4.\quad  Nucleation of massive particles}

\smallskip 

Instead of trying to fix the $S_1\times S_2$ minisuperspace for black
hole nucleation, we shall consider a related problem of nucleation of
massive particles.  The corresponding instanton is the de Sitter $S_4$
with a particle world line in the form of a big circle \cite{14}.  A
similar instanton for nucleation of small extremal black holes was
discussed by Mellor and Moss \cite{2}.  The instanton action is $S_p
=S_0 +2\pi H^{-1}m$, and the nucleation rate is
\begin{equation}
\Gamma\propto\exp[-S_p +S_0]=\exp (-2\pi m/H).
\label{9}
\end{equation}
Particle nucleation in deSitter space has all relevant features of
black hole nucleation, and the Bousso-Hawking argument can be applied
to it without change.  If the argument were true, then the tunneling
wave function would predict a catastrophic instability of de Sitter
space with respect to pair production.  However, we will show that in
fact it gives the correct particle production rate (\ref{9}).

We shall assume that the mass of the created particles is sufficiently
small that their gravitational backreaction can be ignored, $m<<1$, yet
sufficiently large that the concept of particle can be unambiguosly 
defined in our expanding background. As we shall see, this requires 
$m>>H$. This is actually the same mass range in which 
the world line instanton approximation is valid.

Within this regime, we can represent the particles as excitations of
a massive scalar field $\phi$. Let us expand the field configuration as
$\phi=\sum f_n Q_n$, where $Q_n$ are the harmonics on the 3-sphere.
Then the Wheeler-De Witt equation (\ref{8}) can
be solved perturbatively in $f_n$  
using the WKB ansatz 
$\psi\sim \exp iS$, 
where \cite{15}
\begin{equation}
S=S_0(a)+{1\over 2}\sum S_n(a) f_n^2,
\label{9'}
\end{equation}
and
\begin{equation}
{dS_0\over da}=\pm a (a^2H^2-1)^{1/2}.\label{10}
\end{equation}
The solution of the Hamilton-Jacobi equation (\ref{10})
is the action along the classical de Sitter solution,
which represents an inflationary universe. For $a<H^{-1}$ classical motion is 
forbidden, $S_0$ is imaginary, and $\psi$ is a linear combination of growing 
and decaying exponentials. For $a>H^{-1}$ classical motion is allowed and 
$\psi$ is a linear combination of incoming and outgoing waves, whose flux
points towards and away from the forbidden region respectively.
The tunneling boundary condition at $a\to\infty$ 
requires that only the outgoing wave should
be present. The matching conditions for $\psi$ then imply that the 
growing and decaying components have comparable magnitude near the 
classical turning point $a=H^{-1}$. 

With our ansatz, the Wheeler-De Witt equation 
reduces to the functional Schrodinger equation for the quantum field 
$\phi$ in a fixed de Sitter background, with the scale factor playing 
the role of time. Its solution is given by $S_n=H^2 a^2 \dot \nu_n/\nu_n$,
where a dot indicates derivative with respect to the conformal time $\tau$,
defined by $a=(H \cos \tau)^{-1}$, and $\nu_n(\tau)$ are normal modes of the 
classical scalar field equation. 
For each $n$ there are  
two independent solutions, and the quantum state is specified once we 
choose a particular
linear combination of these as our `negative frequency' mode $\nu_n$.
The mode functions $\nu_n$ should in principle be determined by the
outgoing-wave and regularity conditions at the boundaries of
superspace: $a\to 0, ~f_n\to\pm\infty$.  This, however, cannot be
implemented within the perturbative approach, since the expansion
(\ref{9'}) breaks down at large values of $f_n$.  Instead, we shall
follow Ref. \cite{16} and require that the wave function does not
increase towards large $f_n$.  It appears that this is the best one
can do to represent the boundary condition
$\psi(|f_n|\to\infty)<\infty$.  Mathematically, our condition requires that
$Im(S_n)>0$ along 
the three branches of the semiclassical wave function
(growing, decaying and outgoing). It can
be shown that this requirement uniquely determines $\nu_n$, and 
hence the quantum state of the scalar field \cite{16}. 
The corresponding eigenmodes are given by
\begin{equation}
\nu_n \propto cos \tau e^{in\tau}F(\mu,1-\mu,n+1;(1+i \tan \tau)/2),
\label{hyp}
\end{equation}
$$
\mu={1\over 2}-\left({9\over 4}-{m^2\over H^2}\right)^{1/2},
$$
where $\tau$ ranges from $0$ to $\pi/2$ for the outgoing branch and
from $0$ to $\pm i\infty$ for the growing and decaying exponentials.
These modes correspond to the Bunch-Davies or de Sitter invariant
vacuum.  

The Hartle-Hawking wave function for this model includes only the
growing exponential for $a<H^{-1}$ and a linear combination of
incoming and outgoing waves with equal amplitudes at $a>H^{-1}$.  The
mode functions $\nu_n$ are given by Eq. (\ref{hyp}) and are the same
as for the tunneling wave function.  It is easily seen that 
the two wave functions are not related  
by the transformation $S\to -S$, as suggested by Bousso and
Hawking [see Eqs. (\ref{5}), (\ref{6})].

It is well known that a particle detector responds in the Bunch-Davies
vacuum as if there was a thermal distribution of particles with temperature
$H/2\pi$ \cite{18}. This is already in qualitative agreement with the 
instanton result (\ref{9}). The correspondence can be made even more
precise by using the 
method of Bogoliubov coefficients \cite{20}.

For $m>>H$ and for each $n$ an unambiguous definition
of particles can be given at late times. The scalar wave equation is given by
\begin{equation}
\ddot \nu_n+3H\tanh (Ht)\ \dot \nu_n+ m^2 \nu_n + 
{(n^2-1) \over a^2}\nu_n =0,
\label{11}
\end{equation}
where now the dots indicate derivative with respect to cosmological time 
defined by $a=\cosh(Ht)$. For $t>>t_n$ where $t_n$ is the time at which
the particles become non-relativistic,
$ a(t_n)\sim {nH/m}$,
the last term can be ignored and we have approximate solutions of the form
$$
\nu_n \propto a^{-3/2}(\alpha_n e^{i\omega t}+ \beta_n e^{-i\omega t}),
$$
where $\omega\equiv H |\mu-1/2| \approx m$. For large $m$, the exponentials
oscillate fast compared with the expansion rate and we have a good definition  
of positive and negative frequency modes corresponding to our `out' vacuum. 
The Bogoliubov coefficients $\beta_n$ can be readily found by using the 
expansion 
of the hypergeometric functions (\ref{hyp}) at late times 
and the normalization 
condition $|\alpha_n|^2-|\beta_n|^2=1$. The expectation value of the number of
`out' particles in a given mode is then
$$
N_n=|\beta_n|^2=(e^{2\pi \omega/H}-1)^{-1}\approx e^{-2\pi m/H}.
$$
The distribution is independent of $n$, as expected for
nonrelativistic particles. At any given time, the ocupation numbers of
the relativistic modes are exponentially suppressed with respect to
the non-relativistic ones, so the distribution is cut-off at $n\sim
am$. The resulting particle production rate per unit physical volume
is $(dN/dtdV)\sim m^3 H \exp{(-2\pi m/H)}$ \cite{20}. 
In Ref.\cite{19} it was shown that 
the same rate can be obtained by considering the one-loop prefactor to
the instanton contribution. 
There, the cut-off arises because
particles with $n>>a(t)m$ have not been created yet at time $t$.

Hence, the tunneling boundary 
condition does not lead to disastrous pair production, but to a result
which is in good agreement with the instanton calculation.  

In following this tunneling {\it vs.} Hartle-Hawking debate, the
reader should be aware that both wave functions are far from being
rigorous mathematical objects with clearly specified
calculational procedures.  Except in the simplest models, the actual
calculations of $\psi_T$ and $\psi_{HH}$ involve additional
assumptions which appear reasonable, but are not really well
justified.  (The results presented in this paper are no exception).  
For a recent discussion of the problems associated with defining and
interpreting the cosmological wave function see, e.g., Ref. \cite{7'}. 

\bigskip
\noindent{\bf Acknowledgements}

We are grateful to Serge Winitzki for helpful discussions.
This work was partially supported by NATO under grant CRG 951301,
by the U.S. Department of Energy (D.O.E.)
under cooperative research agreement DE-FC02-94ER40818, and by the
National Science Foundation.


\begin{thebibliography}{11}
\bibitem{11}Slices of a 4-geometry ${\cal M}$ can be obtained as
surfaces $f(x)=const$, where $f(x)$ is a smooth real function on
${\cal M}$.  These slices will have smooth geometries, except the
slices passing through critical points where $\partial_\mu f=0$.  The
latter slices have singular geometries.  A singular slice is
non-degenerate if $\det [\partial_\mu\partial_\nu f(x)]\not= 0$ at the
corresponding critical point(s).
\bibitem{12}The corresponding slicing of $S_2\times S_2$ can be
obtained by choosing $f(x)=cos\theta$, where $\theta$ is 
the polar angle on one
of the $S_2$.  Then, $\det [\partial_\mu\partial_\nu f(x)]= 0$
everywhere, and
the singular slices must be degenerate  (see the preceeding footnote
\cite{11}).
\bibitem{20}The Hartle-Hawking wave function gives the same particle 
production
rate, since it predicts the same quantum state of the scalar field.
\end{thebibliography}
\end{document}